\documentclass[linenumbers]{aa}
\usepackage{graphicx,lipsum}
\usepackage[percent]{overpic}

\usepackage{amssymb,amsmath}
\usepackage[version=4]{mhchem}
\usepackage{newtxtext}
\usepackage[varvw]{newtxmath}
\usepackage{xcolor}
\usepackage{chemfig} 
\usepackage{hyperref}

\newcommand{\ignore}[1]{}

\def\aap {{A\&A}}

\def\apjl {{ApJL}}

\def\apjs {{ApJS}}

\begin{document}

   \title{A theory of transmission spectroscopy of hydrodynamic outflows from planetary atmospheres:
Spectral-line saturation and limits on mass-loss constraints}
  \titlerunning{A theory of transmission spectroscopy of hydrodynamic outflows}

\author{
Leonardos Gkouvelis\inst{1}
\thanks{Corresponding author: \email{gkouvelis@iaa.es}}
}

\institute{
Instituto de Astrofísica de Andalucía (IAA-CSIC), 
Glorieta de la Astronomía s/n, E-18008 Granada, Spain
}

    \authorrunning{ Gkouvelis L. }

   \date{Received \today}

 \abstract{
Transmission spectroscopy is a key technique in the characterization of exoplanet atmospheres and has been widely applied to planets undergoing hydrodynamic escape. While a robust analytic theory exists for transmission spectra of hydrostatic atmospheres, the corresponding interpretation for escaping atmospheres has so far relied on numerical modeling, despite the growing number of observations of planetary winds. In this work, a theory of transmission spectroscopy in hydrodynamically escaping atmospheres is developed by coupling the standard transmission geometry to a steady-state, spherically symmetric, isothermal outflow. This approach yields closed-form expressions for the chord optical depth and effective transit radius of a planetary wind and allows the optical depth inversion problem to be examined.The analytic solution reveals that transmission spectroscopy of planetary winds naturally separates into two regimes. In an opacity-limited regime, transmission depths retain sensitivity to the atmospheric mass-loss rate $\dot{M}$. Beyond a critical threshold, however, spectral-line cores become saturated and no longer provide a unique constraint on the escape rate. This transition is marked by a sharp analytic boundary of the form $\sigma(\lambda)\,\dot{M} \le C_{\rm sat}$, where $\sigma(\lambda)$ is the line absorption cross-section and $C_{\rm sat}$ is a constant set by the thermodynamic and geometric properties of the wind. This condition specifies when the inversion between transmission depth and mass-loss rate admits a real solution. Once it is violated, the effective transit radius is no longer controlled by opacity or mass loss, but by the geometric extent of the absorbing wind. These results demonstrate that spectral-line saturation in transmission spectroscopy corresponds to a fundamental loss of invertibility between absorption and atmospheric mass loss, rather than a gradual weakening of sensitivity. The theory provides a physically transparent explanation for why strong transmission line cores, such as the He triplet or Ly$\alpha$, may lose unique sensitivity to mass-loss rates once they enter the saturation
regime, while weaker lines and the wings of strong lines can remain
diagnostic when observationally accessible.}


 \maketitle
 
\section{Introduction}\label{sec:intro}

One of the key techniques on which exoplanet atmospheric characterization relies is
transmission spectroscopy. During a primary transit, a fraction of the
stellar radiation passes through the planetary atmosphere, where it is selectively
absorbed or scattered depending on wavelength and atmospheric composition. As a
result, the effective planetary radius inferred from transit observations varies
with wavelength. This wavelength-dependent modulation of the transit depth encodes
information about atmospheric composition, structure, and opacity sources
\citep{Brown2001}. Currently,  transmission spectra are routinely obtained
and interpreted for a wide range of exoplanets, from hot Jupiters to smaller and
cooler planets, using both space- and ground-based facilities (e.g.,
\citealt{Espinoza2024, Steinrueck2025}).
    
While transmission theory in a broader context was developed early in Earth and
planetary sciences, its application to the observational geometry of
transiting exoplanets was first presented by \citet{Seager2000}.  That work identified that atmospheric opacity controls the apparent transit radius
and provided a qualitative scaling linking opacity, atmospheric scale height, and transit depth. Later, a closed-form analytic solution was
derived by \citet{Lecavelier2008}, who showed that, under the assumptions of a
hydrostatic and isothermal atmosphere, the effective planetary radius scales
logarithmically with opacity as $R_{\mathrm{p}}(\lambda) \sim H \ln \kappa(\lambda)$
where $H$ is the atmospheric scale height and $\kappa(\lambda)$ is the
wavelength-dependent opacity. This logarithmic scaling has served as a canonical
result for more than two decades and has provided valuable intuition for the
interpretation of transmission spectra. More recently, \citet{Gkouvelis2026} derived
a generalized closed-form expression that accounts for pressure-dependent opacity,
modifying the classical scaling to
$R_{\mathrm{p}}(\lambda) \sim \frac{H}{1+n(\lambda)} \ln \kappa_0(\lambda)$, where $n$ is the power-law exponent describing the pressure dependence of the opacity and $\kappa_0$ is a reference opacity. All analytic transmission theories of this kind
rely on a common set of assumptions, namely hydrostatic equilibrium, an isothermal
atmosphere, and the absence of strong compositional gradients along the transit
chord. In addition, well known degeneracies affect the normalization of hydrostatic
transmission spectra, limiting the unique retrieval of absolute atmospheric
properties \citep{Benneke2012,deWit2013}. Together, these results provide a robust
theoretical framework for building intuition and interpreting transmission spectra
of stable, hydrostatic atmospheres.

Nevertheless, a significant fraction of the known exoplanet population resides in a non-hydrostatic regime, commonly referred to as a planetary wind \citep{Watson1981,Owen2019}, whose onset and long-term persistence can strongly influence the volatile inventory, atmospheric evolution, and potential habitability of rocky planets \citep{Gkouvelis2025}. Under conditions of rapid atmospheric escape, the upper
atmosphere is no longer in hydrostatic balance but instead undergoes hydrodynamic
outflow, continuously losing mass to space. Transmission spectroscopy has been the
primary observational tool for studying this process, mainly through targeted
spectral lines with large absorption cross sections that probe atmospheric layers
where the outflow is established and, in some cases, extends beyond the Roche lobe
(e.g., Ly$\alpha$, He~\textsc{i}~1083~nm; \citealt{Bourrier2016, Oklopvcic2018}).

To date, the interpretation of these observations has relied on numerical models of
hydrodynamic escape. One recurring result of such studies is that the cores of strong
spectral lines rapidly become insensitive to the mass-loss rate, such that they
primarily encode the geometric extent of the absorbing atmosphere rather than the
magnitude of the mass flux
(\citealt{Allan2019, Linssen2022, DosSantos2022}).
Recent work has highlighted that strong transmission lines, particularly
Ly$\alpha$, do not primarily probe atmospheric mass-loss rates but are instead
controlled by geometric and ionization constraints set by the outflow and the
stellar environment \citep{Owen2023}.

In contrast, information about the hydrodynamic flow is typically carried by the
wings of spectral lines, which remain sensitive to the atmospheric column density
and velocity structure
(\citealt{BallabioOwen2025, Lampon2023}). This distinction has so far emerged empirically from numerical modeling, rather than from an analytic transmission framework. These results suggest that the information content of wind transmission spectra is not distributed uniformly across a line profile.

In this work, the standard transmission geometry is coupled to an isothermal, spherically symmetric, steady-state hydrodynamic outflow, and a closed-form expression for the effective transit radius of planetary winds is derived. For wavelengths
that probe the atmospheric flow, the analytic solution provides a useful
approximation to the transmission spectrum and offers physical insight into its
dependence on opacity, temperature, and mass-loss rate. The
analytic framework reveals a fundamental limitation of transmission spectroscopy
in escaping atmospheres. It is shown that the mapping between transmission depth and
atmospheric mass loss exists only over a restricted range of parameters. Beyond a
critical threshold, strong spectral lines become saturated in such a way that the
inversion between absorption and mass loss is no longer unique. In this regime, the
effective transit radius is set primarily by the geometric extent of the absorbing
wind rather than by the mass flux itself. This behavior provides a physical explanation for why the cores of strong transmission lines often lose unique sensitivity to mass-loss
rates, while diagnostic information about the hydrodynamic flow may be
retained in weaker lines and in the wings of strong transitions.
The present theory is intended to identify this intrinsic
radiative-transfer limitation. In practice, however, the interpretation
of individual tracers may also be affected by additional observational
and thermochemical degeneracies, such as interstellar absorption,
uncertain level populations, and temperature-dependent ionization or
excitation structure.

This work is organized as follows. In Sect.~2, the hydrodynamic outflow model is described and analytic expressions for the chord optical depth and effective transit radius in a steady-state planetary wind are derived. In Sect.~3, the properties and domain of validity of the analytic solution are analyzed, and the regimes in which the inversion between transmission depth and mass-loss rate breaks down are identified. In Sect.~4, the analytic predictions are compared to numerical transmission spectra and the physical interpretation of saturated and opacity-limited regimes is discussed. Finally, Sect.~5 summarizes the main results and their implications for the interpretation of transmission observations of escaping exoplanet atmospheres.

\section{Analytic derivation} \label{sec:derivation}

\subsection{Physical assumptions}

\begin{figure}[ht!]
    \centering
    \includegraphics[scale=0.27, trim=0.cm 1.5cm 3cm 0cm, clip]{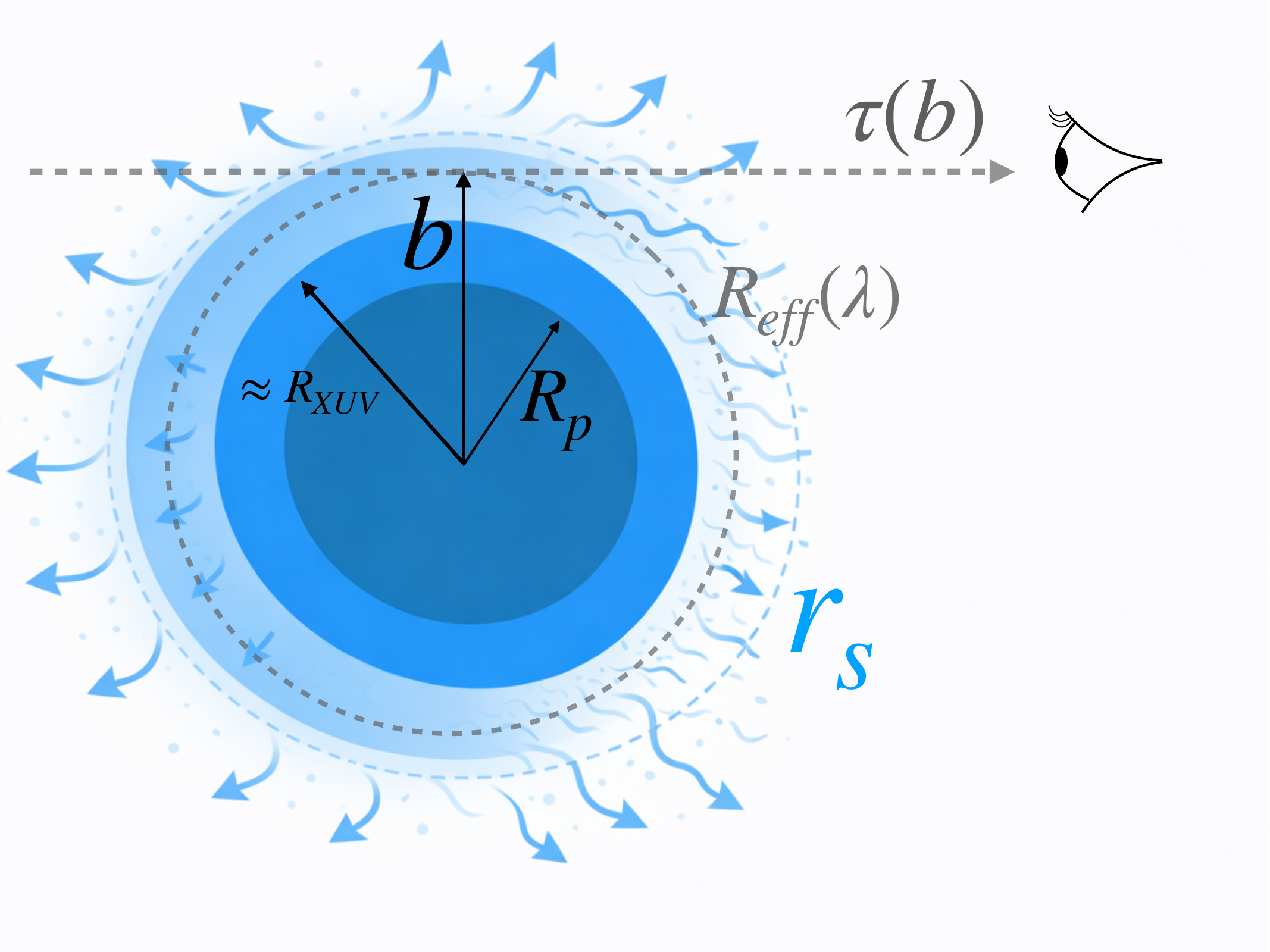}
    \caption{
Geometry of transmission spectroscopy in a planetary wind.
A planet of radius $R_p$ is surrounded by a bound hydrostatic atmosphere and an outer hydrodynamically escaping region.
Stellar rays intersect the atmosphere along chords of impact parameter $b$, accumulating a wavelength-dependent slant optical depth $\tau(b)$ along the line of sight toward the observer.
The effective transit radius $R_{\mathrm{eff}}(\lambda)$ is defined by the impact parameter for which $\tau(b)$ reaches the reference value $\tau_\ast$.
The dashed  blue circle marks the sonic radius $r_s$ and
the approximate location of the XUV photosphere, $R_{\mathrm{XUV}}$, is also indicated.
}

    \label{fig:cartoon}
\end{figure}

We examine the planetary winds assuming a steady-state,  spherically symmetric and isothermal flow, which has the  mathematical formulation of the Parker wind solution for the Solar wind (\citet{Parker1958, Lamers1999}).

Combining the momentum equation, the mass conservation and the isothermal equation of state, the  Parker wind equation is obtained
\begin{equation}
    \left( v^2 - c_s^2 \right)\frac{1}{v}\frac{dv}{dr}
    = \frac{2 c_s^2}{r} - \frac{G M_p}{r^2}.
    \label{eq:parker_diff}
\end{equation}

The \emph{sonic radius} $r_s$ is defined by the condition $v(r_s) = c_s$. Requiring Eq.~\eqref{eq:parker_diff} to be
regular at $r = r_s$ gives
\begin{equation}
    r_s = \frac{G M_p}{2 c_s^2}.
    \label{eq:r_sonic}
\end{equation}

The transonic Parker solution can be written in implicit form as
\begin{equation}
    \left( \frac{v}{c_s} \right)^2
    - \ln\!\left[ \left( \frac{v}{c_s} \right)^2 \right]
    = 4 \ln\!\left( \frac{r}{r_s} \right) + 4 \frac{r_s}{r} - 3.
    \label{eq:parker_implicit}
\end{equation}

Solving  for $n(r)$  in the mass conservation equation we have the density profile in the radial axis:
\begin{equation}
    n(r) = n_s \left( \frac{r_s}{r} \right)^2 \frac{c_s}{v(r)}.
    \label{eq:n_of_r_general}
\end{equation}

\subsection{Transit geometry}
I now consider a ray of starlight passing at an impact parameter $b$ (measured from the planetary center). Along the line-of-sight coordinate $x$, the radial coordinate is
\begin{equation}
    r^2 = x^2 + b^2,
\end{equation}

so that the monochromatic chord optical depth at wavelength $\lambda$ is 

\begin{equation}
    \tau(b,\lambda)
    = 2 \int_{r=b}^{+\infty} \sigma(\lambda)\, n(r)\,
      \frac{r}{\sqrt{r^2 - b^2}}\, dr.
    \label{eq:tau_general}
\end{equation}

The transmission geometry and the main characteristic radii of the flow are illustrated in Fig.~\ref{fig:cartoon}.
Substituting Eq.~\eqref{eq:n_of_r_general} into the optical depth gives
\begin{equation}
    \tau(b,\lambda)
    = 2 \sigma(\lambda)\, n_s\, c_s\, r_s^2
      \int_{b}^{+\infty}
      \frac{dr}{v(r)\, r\, \sqrt{r^2 - b^2}}.
    \label{eq:tau_parker_exact}
\end{equation}

This is the exact chord optical depth in a Parker wind, expressed in terms of the velocity profile $v(r)$ and the sonic-point parameters $(r_s, n_s)$.

To retain analytical tractability, the integral is approximated in the subsonic region of the Parker wind. The geometric kernel $1/\sqrt{r^2-b^2}$ peaks strongly near the tangent point $r = b$,
so the main contribution to the integral comes from a narrow region around $r \approx b$. 
This approximation is analogous to the tangent-point treatment commonly adopted
in analytic transmission spectroscopy, where the chord optical depth is dominated
by a narrow region around the point of closest approach and slowly varying
quantities may be evaluated locally \citep[e.g.][]{Lecavelier2008, deWit2013}.
In this region, the wind speed varies slowly compared to the geometric factor, and I
can approximate $v(r) \approx v(b) $. With this approximation, $v(b)$ is constant with respect to the integration variable $r$,  so that I can factor it out, Eq.~\eqref{eq:tau_parker_exact} becomes

\begin{equation}
\tau(b,\lambda)
\simeq \frac{2\,\sigma(\lambda)\,n_s\,c_s\,r_s^2}{v(b)}
\int_{b}^{+\infty} \frac{dr}{r\,\sqrt{r^2 - b^2}} .
\label{eq:tau_after_vb}
\end{equation}

The remaining integral is purely geometrical and can be evaluated analytically. One finds
\begin{equation}
\int_{b}^{+\infty} \frac{dr}{r\,\sqrt{r^2 - b^2}} = \frac{\pi}{2b}.
\label{eq:geom_integral}
\end{equation}
Inserting Eq.~\eqref{eq:geom_integral} into Eq.~\eqref{eq:tau_after_vb}, I obtain
\begin{equation}
\tau(b,\lambda)
\simeq \frac{2\,\sigma(\lambda)\,n_s\,c_s\,r_s^2}{v(b)}\,
\frac{\pi}{2b}
= \frac{\pi\,\sigma(\lambda)\,n_s\,c_s\,r_s^2}{b\,v(b)} .
\label{eq:tau_parker_general}
\end{equation}
Equation~\eqref{eq:tau_parker_general} is a general expression for the chord optical depth
in a Parker wind, valid in the approximation that the wind speed is nearly constant
over the narrow region around the tangent point $r \approx b$.

\subsection{Subsonic approximation}
\begin{figure}[ht!]
    \centering
    \includegraphics[scale=0.54, trim=0cm 0cm 0cm 0cm, clip]{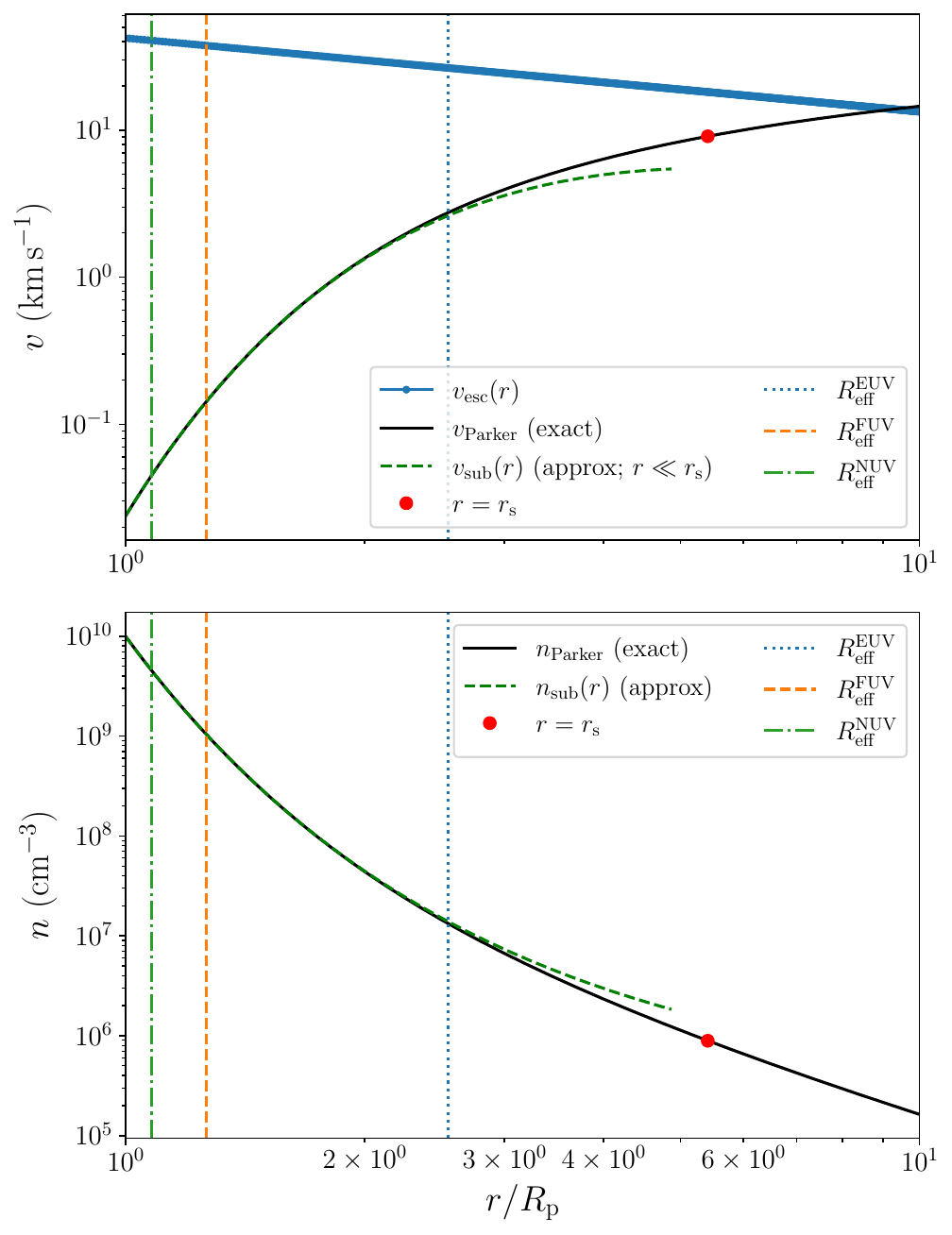}
    \caption{ Radial velocity and density profiles of the Parker wind for the hot Jupiter HD~209458\,b. With black solid lines I show the profiles as calculated by the Parker wind formulation while in dashed green the subsonic approximation (see Section~\ref{sec:derivation}). For comparison, I overplot the radii where the optical depth of the NUV, FUV, EUV are reaching unity, $\tau\approx 1$.
}
    \label{fig:parker}
\end{figure}

In the subsonic region $r \ll r_s$, the isothermal Parker wind admits a simple
asymptotic expression for the velocity (see, e.g., \citealt{Parker1958}).
Starting from the implicit relation \eqref{eq:parker_implicit}, one finds that for
$v \ll c_s$ the velocity can be written as
\begin{equation}
    v(r) \simeq c_s\,e^{3/2}\,\left( \frac{r_s}{r} \right)^2
    \exp\!\left( - 2\frac{r_s}{r} \right),
    \label{eq:parker_subsonic_v}
\end{equation}
which is valid for radii well below the sonic point, $r \ll r_s$. 
The numerical prefactor in Eq.~\ref{eq:parker_subsonic_v} depends on the normalization of the transonic solution and is accurate up to a factor of order unity, which does not affect the scaling relations or the qualitative results derived below. A comparison between the exact Parker solution and the subsonic asymptotic approximation is shown in Fig.~\ref{fig:parker}.

Evaluating Eq.~\eqref{eq:parker_subsonic_v} at the tangent point $r = b$ and substituting Eq.~\eqref{eq:parker_subsonic_v} into Eq.~\eqref{eq:tau_parker_general} yields
\begin{equation}
    \tau(b,\lambda)
    \simeq \frac{\pi\,\sigma(\lambda)\,n_s\,c_s\,r_s^2}{b\,v(b)}
    = \pi\,\sigma(\lambda)\,n_s\,e^{-3/2}\,
    b\,\exp\!\left( 2\frac{r_s}{b} \right).
    \label{eq:tau_subsonic_prefactor}
\end{equation}
For later convenience, a wavelength-dependent prefactor is defined
\begin{equation}
    A(\lambda) \equiv \pi\,\sigma(\lambda)\,n_s\,e^{-3/2},
    \label{eq:A_def_ns}
\end{equation}
so that Eq.~\eqref{eq:tau_subsonic_prefactor} can be written compactly as
\begin{equation}
    \tau(b,\lambda) \simeq A(\lambda)\,b\,
    \exp\!\left( 2\frac{r_s}{b} \right),
    \qquad (b \ll r_s).
    \label{eq:tau_subsonic_simple}
\end{equation}

\begin{figure}[ht!]
    \centering
    \includegraphics[scale=0.54, trim=0.25cm 0cm 0cm 0cm, clip]{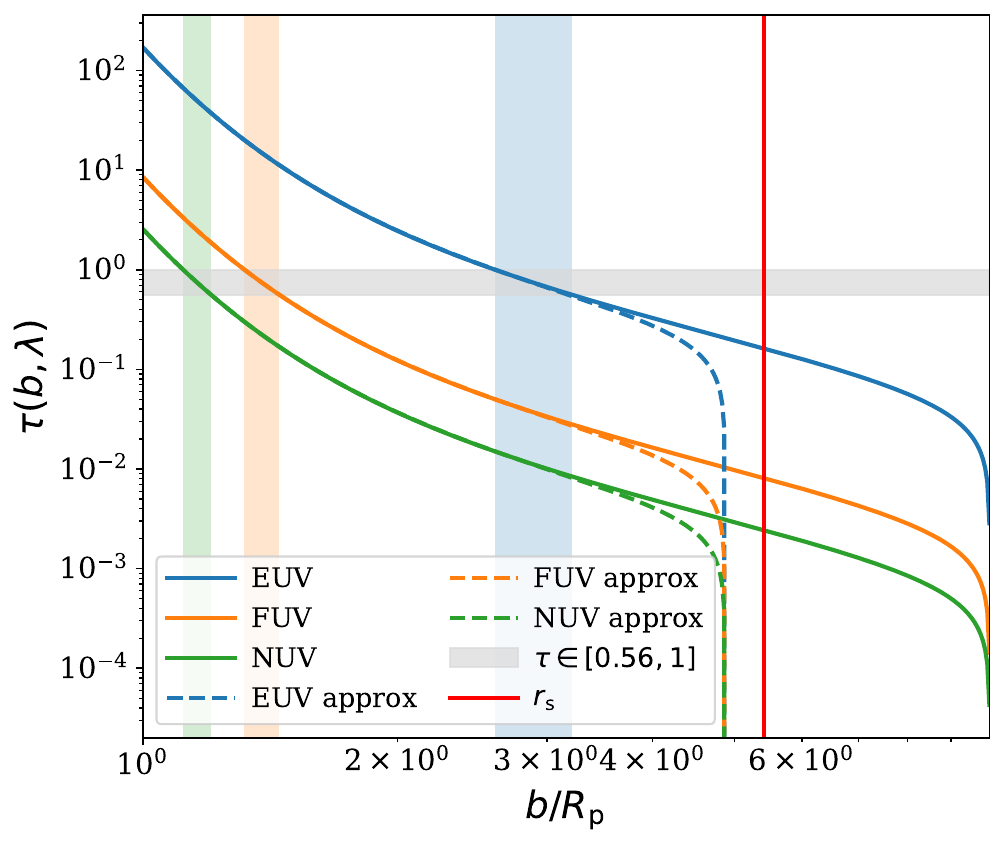}
    \caption{Optical depth as a function of planet radius for EUV, FUV and NUV  wavelength bands as well as approximations for the same bands overplotted. With shaded stripes I show the range $\tau=0.56$-1 expressed in terms of the corresponding planet radii for the ultraviolet bands shown.
}
    \label{fig:tau}
\end{figure}

An example of the resulting chord optical depth profiles for different ultraviolet bands is shown in Fig.~\ref{fig:tau}.

The sonic density is then related to the mass-loss rate. With my assumptions the mass loss rate is written as

\begin{equation}
    \dot{M} = 4\pi r^2 \rho(r)\,v(r),
\end{equation}
Solving  for $n_s$ gives and substituting into Eq.~\eqref{eq:A_def_ns}, the prefactor
$A(\lambda)$ can be written directly in terms of the mass-loss rate:
\begin{equation}
    A(\lambda)
    = \pi\,\sigma(\lambda)\,e^{-3/2}\,
      \frac{\dot{M}}{4\pi \mu m_p c_s r_s^2}
    = \frac{\sigma(\lambda)\,\dot{M}}{4\,e^{3/2}\,\mu m_p c_s r_s^2}.
    \label{eq:A_def_mdot}
\end{equation}

\subsection{Effective transit radius and Lambert-W solution}
\label{sec:Reff_Lambert}

In the standard definition, the effective transit radius $R_{\rm eff}(\lambda)$ is obtained from
the total obscured  area,
\begin{equation}
    R_{\rm eff}^2(\lambda)
    = R_0^2 + 2 \int_{R_0}^{+\infty}
      \bigl[ 1 - e^{-\tau(b,\lambda)} \bigr]\,b\,db,
    \label{eq:Reff_def}
\end{equation}
where $R_0$ is a reference radius (\citet{Lecavelier2008, Gkouvelis2026}).
When $\tau(b,\lambda)$ is a steep function of $b$, the transmission can be approximated by a sharp transition in opacity.  This step-function approximation is standard in  transmission theory when the chord optical depth varies rapidly with impact parameter \citep[e.g.][]{Lecavelier2008, Brown2001}. In this regime it is a good approximation to treat the transmission as
a sharp transition at an impact parameter $b_\ast(\lambda)$ where the chord optical depth
reaches a reference value $\tau_\ast \sim \mathcal{O}(1)$ (common choices are
$\tau_\ast = 1$ or $\tau_\ast \simeq 0.56$ \citep[e.g.][]{Lecavelier2008}). To leading order in this step-function
approximation the effective radius is $R_{\rm eff}(\lambda) \approx b_\ast(\lambda)$, where $b_\ast$ is defined implicitly by $ \tau\bigl(b_\ast,\lambda\bigr) = \tau_\ast$.  
This steep-$\tau$ approximation provides a closed analytic solution
that is valid when the transition between optically thin and optically
thick regions occurs over a narrow range of impact parameters. It
therefore defines a local inversion between transmission depth and
mass-loss rate in the regime where the effective transit radius can be
associated with a characteristic optical-depth surface. The physical
interpretation of the resulting saturation boundary in the case of more gradually varying
$\tau(b)$, is discussed in Sect.~\ref{sec:hydro_interp}.

Combining this condition with the subsonic Parker expression
(Eq.~\eqref{eq:tau_subsonic_simple}), I obtain
\begin{equation}
\tau_\ast
= A(\lambda)\,b_\ast\,
\exp\!\left( 2\frac{r_s}{b_\ast} \right),
\label{eq:bstar_eq_A_compact}
\end{equation}
or equivalently
\begin{equation}
D(\lambda)
\equiv \frac{\tau_\ast}{A(\lambda)}
= b_\ast\,
\exp\!\left( 2\frac{r_s}{b_\ast} \right),
\label{eq:D_bstar}
\end{equation}
where $D(\lambda)$ has dimensions of length. Finally, I can solve for $b_\ast$ in closed form using the Lambert-$W$
function (\citet{Corless1996}).

I define $y \equiv -2r_s/b_\ast$, so that $b_\ast = -2r_s/y$. Substituting into
Eq.~\eqref{eq:D_bstar} gives $D = (-2r_s/y)e^{-y}$, and rearranging yields
\begin{equation}
y\,e^{y} = -\,\frac{2r_s}{D(\lambda)}.
\label{eq:yW_eq}
\end{equation}

By definition of the Lambert-$W$ function, this implies
\begin{equation}
    y = W\!\left( -\,\frac{2r_s}{D(\lambda)} \right).
    \label{eq:y_W}
\end{equation}
Using $b_\ast = -2r_s/y$, I obtain
\begin{equation}
    b_\ast(\lambda) = -\,\frac{2r_s}{W\!\left( -\,2r_s/D(\lambda) \right)}.
    \label{eq:bstar_final_D}
\end{equation}
To leading order in the step-function approximation, the effective transit radius is
\begin{equation}
    R_{\rm eff}(\lambda)
    \approx b_\ast(\lambda)
    = -\,\frac{2r_s}{W\!\left( -\,2r_s/D(\lambda) \right)}.
    \label{eq:Reff_final_D}
\end{equation}
Using $D(\lambda)=\tau_\ast/A(\lambda)$ and the definition of $A(\lambda)$
(Eq.~\eqref{eq:A_def_mdot}), I obtain

\begin{equation}
\boxed{
    R_{\rm eff}(\lambda)
    \approx
    -\,\frac{2r_s}{
        W\!\left[
        -\,\dfrac{\sigma(\lambda)\,\dot{M}}
                  {2\,e^{3/2}\,\tau_\ast\,\mu m_p c_s r_s}
        \right] }
}
\label{eq:Reff_final_mdot}
\end{equation}

Equation~\eqref{eq:Reff_final_mdot} is a closed-form expression for the effective transit
radius of a steady state isothermal flow in terms of the sonic radius $r_s$, sound speed $c_s$,
mass-loss rate $\dot{M}$, mean molecular weight $\mu$, and the wavelength-dependent
cross section $\sigma(\lambda)$. The  Lambert-$W$  solution branch determines the
physically relevant solution and is discussed in Section~\ref{sec:branches}.

\section{Validity, branches, and saturation boundary}
\label{sec:validity}

The effective transit radius is obtained by inverting the condition 
$\tau(b,\lambda)=\tau_\ast$. This inversion is mathematically equivalent to solving 
a Lambert-$W$ equation. Whether this equation admits a real solution determines 
whether transmission spectroscopy provides a unique mapping between absorption 
depth and atmospheric mass loss. I thus demonstrate that this mapping exists only 
below a sharp threshold in $\sigma(\lambda)\dot{M}$, and derive this threshold 
explicitly.

\subsection{Lambert-$W$ branches and the physical solution}
\label{sec:branches}

The regime of validity of Eq.~\eqref{eq:Reff_final_mdot} is now discussed and the identification of
the Lambert-$W$ branch that yields a physically meaningful solution in the subsonic
region.

The Lambert-$W$ function is defined implicitly by $W(z)\,e^{W(z)}=z$.
For real arguments $z$, the function has a single real branch $W_0(z)$ for $z\ge 0$,
two real branches $W_0(z)$ and $W_{-1}(z)$ for $-1/e \le z < 0$, and only complex values
for $z<-1/e$ \citep{Corless1996}. On the interval $-1/e \le z < 0$, the principal branch
satisfies $-1 \le W_0(z) < 0$, while the lower branch satisfies $W_{-1}(z)\le -1$.

For this application, it is convenient to define
\begin{equation}
    z(\lambda) \equiv
    -\,\frac{\sigma(\lambda)\,\dot{M}}
              {2\,e^{3/2}\,\tau_\ast\,\mu m_p c_s r_s},
    \label{eq:z_def}
\end{equation}
so that Eq.~\eqref{eq:Reff_final_mdot} becomes
\begin{equation}
    R_{\rm eff}(\lambda)
    \approx -\,\frac{2r_s}{W\!\bigl(z(\lambda)\bigr)}.
    \label{eq:Reff_W_z}
\end{equation}
Since $\sigma(\lambda) > 0$ and $\dot{M} > 0$, I have $z(\lambda) < 0$ for all wavelengths.

The derivation leading to Eq.~\eqref{eq:Reff_W_z} assumes that the effective radius lies
deep in the subsonic region of the Parker wind, $R_{\rm eff}(\lambda) \ll r_s$.
Introducing
\begin{equation}
    y \equiv -\,\frac{2r_s}{R_{\rm eff}},
    \label{eq:y_Reff}
\end{equation}
I have $y=W\!\bigl(z(\lambda)\bigr)$ and $R_{\rm eff}=-2r_s/y$.
The subsonic requirement $R_{\rm eff}\ll r_s$ implies $|y|\gg 2$, i.e.\ the solution of
$y\,e^y=z(\lambda)$ must have large negative magnitude. On $-1/e\le z<0$, the two real
branches behave differently:

\noindent (i) On the principal branch $W_0(z)$, $-1\le W_0(z)<0$, hence $|y|\lesssim 1$
and $R_{\rm eff}\gtrsim 2r_s$, inconsistent with $R_{\rm eff}\ll r_s$.

\noindent (ii) On the lower branch $W_{-1}(z)$, $W_{-1}(z)\le -1$ and
$|W_{-1}(z)|\rightarrow \infty$ as $z\rightarrow 0^{-}$, implying $|y|\gg 1$ and thus
$R_{\rm eff}\ll 2r_s$, consistent with the subsonic approximation.

 Therefore, within the regime where the subsonic Parker approximation is valid and
$z(\lambda)\in[-1/e,0)$, the physically relevant solution is obtained by choosing the $W_{-1}$ branch:
\begin{equation}
\boxed{
    R_{\rm eff}(\lambda)
    \approx -\,\frac{2r_s}{W_{-1}\!\bigl(z(\lambda)\bigr)},
    \qquad
    -\frac{1}{e} \le z(\lambda) < 0
}
\label{eq:Reff_W_minus1}
\end{equation}

\subsection{Real versus complex solutions and the onset of saturation}
\label{sec:real_vs_complex}

For $z(\lambda)<-1/e$, the Lambert--$W$ function has no real values, and
Eq.~\eqref{eq:Reff_W_z} yields a complex $R_{\rm eff}$ that has no direct geometric
interpretation. In practice, a complex solution is best interpreted as a diagnostic that
one (or more) assumptions entering the analytic inversion have broken down. In the present
context, the most relevant interpretation is that the optical-depth criterion
$\tau(b,\lambda)=\tau_\ast$ cannot be satisfied at any $b$ within the subsonic regime because
the line is saturated: the chord optical depth exceeds $\tau_\ast$ for all grazing chords that
remain in the region where the subsonic approximation applies.

Equivalently, the real-domain condition $z(\lambda)\ge -1/e$ defines a sharp boundary in
$(\sigma,\dot M)$ space beyond which the analytic inversion ceases to exist as a real-valued mapping.

\subsection{A single dimensionless control parameter and a quantitative validity boundary}
\label{sec:chi_boundary}

\begin{figure}[ht!]
    \centering
    \includegraphics[scale=0.26, trim=1cm 0cm 0cm 0cm, clip]{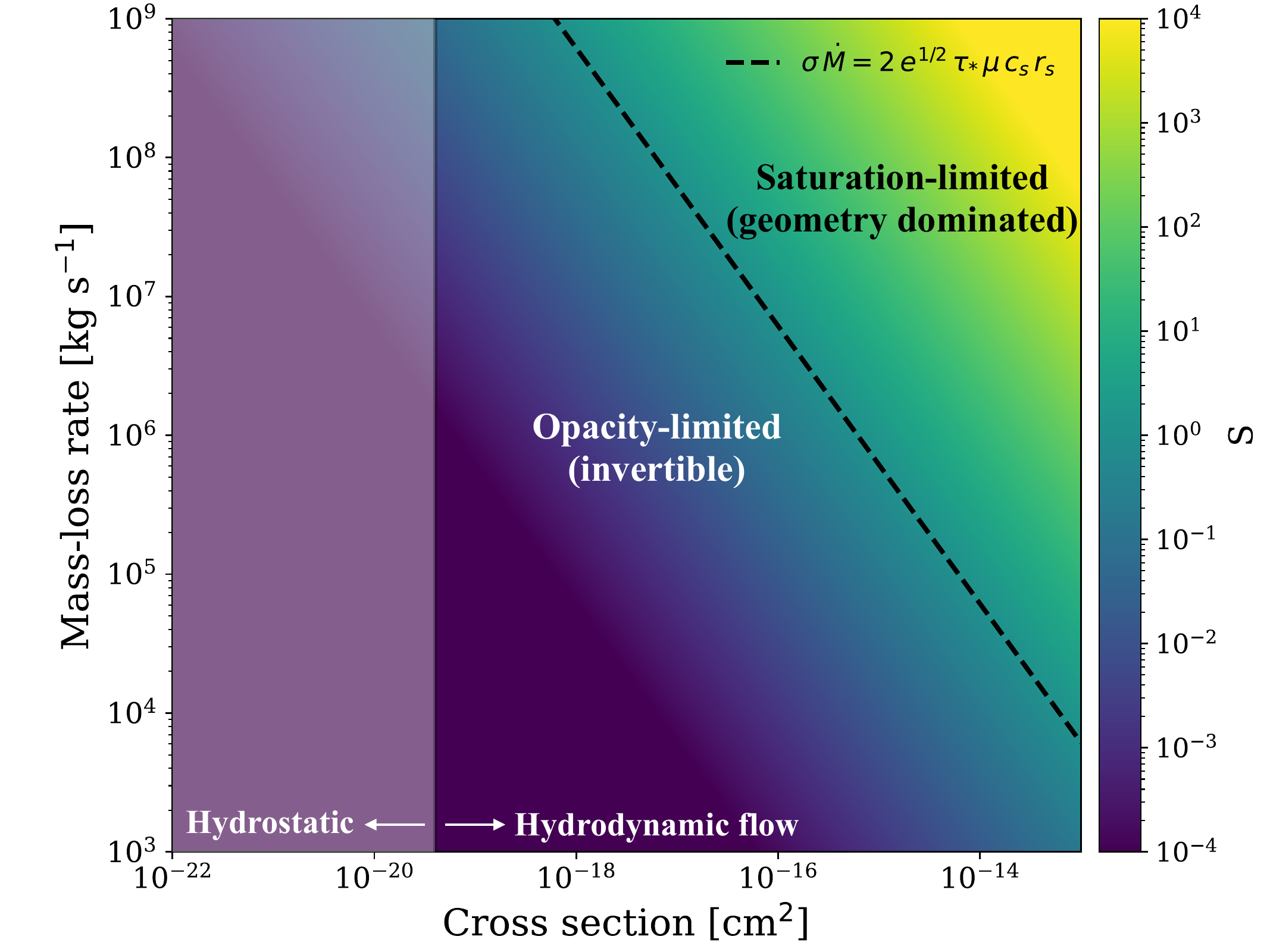}
    \caption{ 
Regime map for transmission through a hydrodynamic planetary wind, shown in the plane
of absorption cross section $\sigma(\lambda)$ and mass-loss rate $\dot{M}$ for
HD~209458\,b. The color scale represents the dimensionless saturation parameter
$S \equiv \sigma(\lambda)\dot{M}/C_{\rm sat}$, where
$C_{\rm sat} =2 e^{1/2}\tau_\ast \mu m_p c_s r_s$.  The dashed black curve marks the analytic validity boundary
$\sigma(\lambda)\dot{M} = C_{\rm sat}$, separating the opacity-limited regime ($S \ll 1$),
in which the transmission depth scales linearly with $\sigma\dot{M}$, from the
saturation-limited regime ($S \gg 1$), in which the effective transit radius is
set primarily by geometry. The light-gray shaded region indicates cross sections
$\sigma \lesssim \sigma_{\rm obs,base}\sim4\times10^{-20}\,\mathrm{cm}^2$, for which
absorption becomes optically thin above the XUV heating base and the observed signal
is expected to probe below the wind launch region. This figure illustrates the
sharp, quantitative boundary between invertible and non-invertible transmission
regimes predicted by the analytic model.
}
    \label{fig:sigmaMassLoss}
\end{figure}

The analytic expression for the effective transit radius in a Parker wind,
Eq.~\eqref{eq:Reff_W_minus1}, depends on wavelength only through the product of opacity and mass-loss rate.
This motivates defining a dimensionless control parameter
\begin{equation}
\chi(\lambda) \equiv \frac{\sigma(\lambda)\,\dot M}{\mu\,m_p\,c_{\rm s}\,r_{\rm s}},
\end{equation}
such that
\begin{equation}
z(\lambda) = -\frac{\chi(\lambda)}{2 e^{3/2}\tau_\ast}.
\end{equation}
The analytic solution exists only for real values on the $W_{-1}$ branch, requiring
\begin{equation}
-\frac{1}{e} \le z(\lambda) < 0,
\end{equation}
which is equivalent to an upper bound on $\sigma(\lambda)\dot M$:
\begin{equation}
\boxed{
\sigma(\lambda)\,\dot M
\;\le\;
C_{\rm sat}
}
\label{eq:validity_boundary}
\end{equation}
where it is defined $C_{\rm sat}\equiv 2e^{1/2}\tau_\ast \mu m_p c_s r_s$. This inequality is the quantitative saturation boundary: when it is satisfied, the inversion between
absorption and effective transit radius is real-valued and well defined; when it is violated, the
corresponding wavelength lies in a saturation-limited regime in which the analytic inversion breaks down. This analytic saturation boundary and the two transmission regimes are illustrated quantitatively in Fig.~\ref{fig:sigmaMassLoss} utilizing HD~209458\,b as an example.

\subsection{Limitations of the asymptotic expansion}
\label{sec:W_asymptotic_limits}

One may formally expand the Lambert--$W$ function for small $|z|$ as
\begin{equation}
    W(z) = z - z^2 + \frac{3}{2}z^3 + \mathcal{O}(z^4),
    \label{eq:W_series}
\end{equation}
which yields a simple approximate scaling of $R_{\rm eff}$ with $\sigma(\lambda)$ and $n_s$.
However, the series \eqref{eq:W_series} is an expansion around $z=0$ on the principal branch $W_0$,
where $W(z)\rightarrow 0$ as $z\rightarrow 0$. Because $W_0(z)\in[-1,0)$ for $z\in[-1/e,0)$, this would imply
$R_{\rm eff}\gtrsim 2r_s$, i.e.\ outside the subsonic region where my Parker asymptotics were derived.

By contrast, on the physically relevant branch $W_{-1}$ one has $W_{-1}(z)\rightarrow -\infty$ as
$z\rightarrow 0^{-}$, and the appropriate asymptotic is logarithmic rather than a power series
\citep[e.g.][]{Corless1996}. Consequently, the small-$|z|$ expansion~\eqref{eq:W_series} is mathematically
correct but not self-consistent for the physical regime of interest ($R_{\rm eff}\ll r_s$). For quantitative
work, it is therefore recommended using Eq.~\eqref{eq:Reff_W_minus1} with the $W_{-1}$ branch and checking
\emph{a posteriori} that $R_{\rm eff}(\lambda)\ll r_s$ for the parameters of interest.

\begin{figure*}[ht!]
    \centering
    \includegraphics[scale=0.5, trim= 0cm 0cm 0cm 0cm, clip]{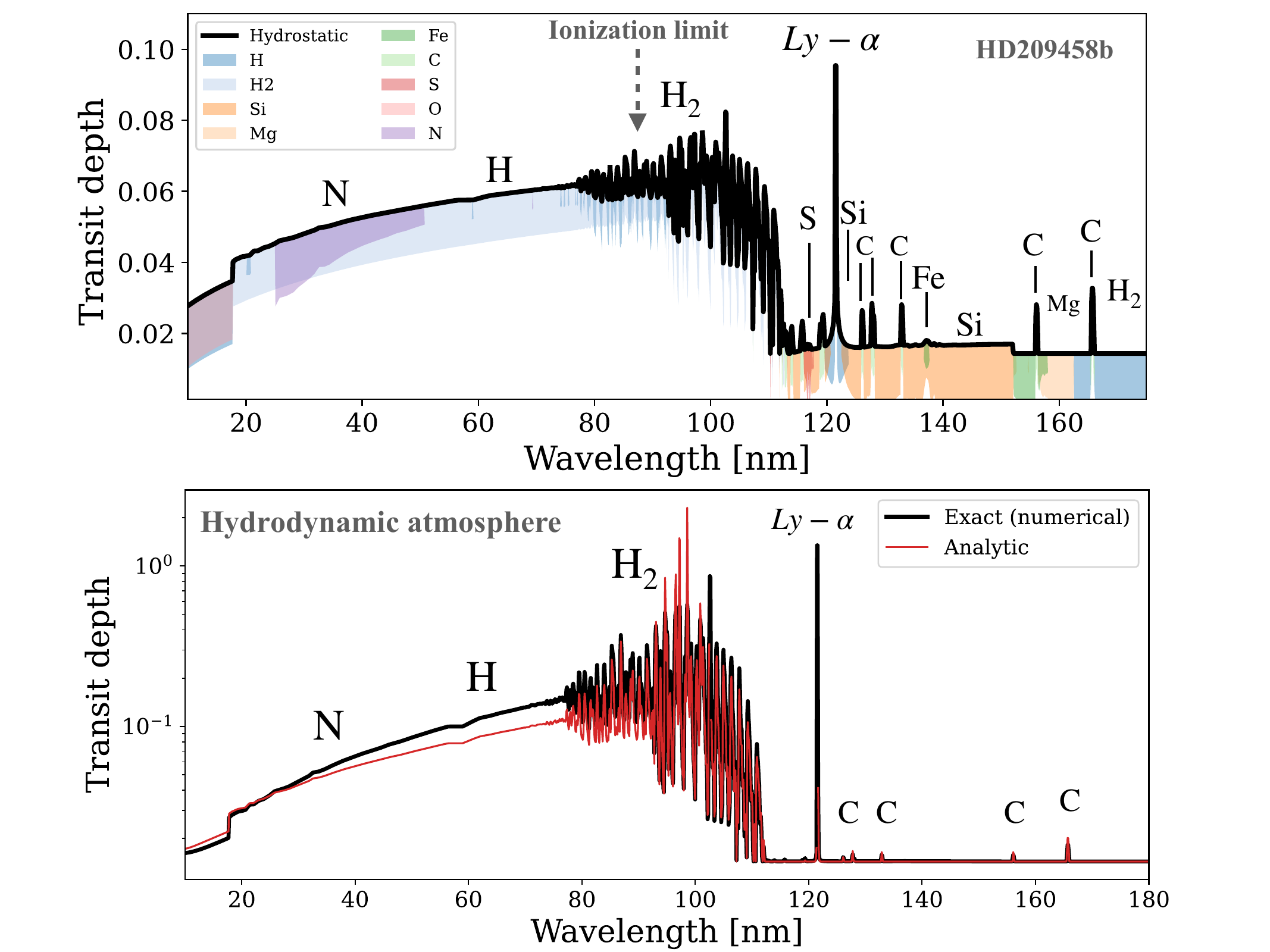}
    \caption{Up: Transmission spectrum of HD209458b under the hydrostatic/stable thermosphere hypothetical scenario. With shaded regions the dominant absorption species at each wavelength are indicated. Bottom: Comparison of HD209458b hydrodynamic atmosphere's transmission spectrum from numerical integration including line broadening effects from bulk flow and thermal motion, compared to the analytic model derived in Eq.~\eqref{eq:Reff_W_minus1}.
}
    \label{fig:synthetic}
\end{figure*}

\section{Physical interpretation and synthetic spectra}
\label{sec:physical_interpretation}

\subsection{Analytic transmission synthetic spectra of planetary winds}
\label{sec:synthetic_spectra}

\paragraph{\textit{Setup and numerical framework.}}
The analytic model for non-hydrostatic atmospheres is now tested by computing synthetic
transmission spectra. To trace the upper atmosphere and thus the hydrodynamic flow, we can focus on
wavelength regions with large absorption cross sections. Ultraviolet wavelengths are particularly
well suited for this purpose, especially for broadband coverage.

In Fig.~\ref{fig:synthetic} a synthetic transmission spectra for the hot Jupiter
HD~209458\,b are presented, a benchmark system that is well studied observationally and known to undergo
hydrodynamic atmospheric escape \citep{Vidal2003, Vidal2004, Murray2009}.
Spectra in the ultraviolet range ($\lambda \approx 15$-180~nm) are computed for comparison between a
hydrostatic (upper panel) and a hydrodynamic Parker-wind scenario (lower panel).
The numerical spectra are computed following Appendix~\ref{app:dnumerical}. Planetary and atmospheric
parameters are adopted from representative literature values (e.g.\ \citealt{Koskinen2010, Koskinen2013}).
Ultraviolet photoabsorption cross sections are compiled from multiple sources
\citep{Heays2017, Gkouvelis2018, Chubb2024, Gkouvelis2024}, and references therein.

\paragraph{\textit{Failure of hydrostatic transmission intuition.}}
The hydrostatic calculation (upper panel of Fig.~\ref{fig:synthetic}) is included only as a baseline
that illustrates what ultraviolet opacities would imply under the assumptions of hydrostatic
transmission theory. The resulting spectral morphology differs qualitatively from the wind case:
once the atmosphere is not in hydrostatic balance, the mapping between opacity, structure, and
transit depth is fundamentally altered. This demonstrates that the physical intuition built from
analytic transmission theory for hydrostatic atmospheres does not carry over to the planetary-wind
regime.

\paragraph{\textit{Analytic--numerical comparison and saturation.}}
In the lower panel of Fig.~\ref{fig:synthetic}, the black curve shows the numerical transmission
spectrum obtained by line-by-line integration through the full Parker wind density profile
(Eq.~\ref{eq:n_of_r_general}). The red curve shows the analytic prediction from
Eq.~\eqref{eq:Reff_W_minus1}. The overall agreement confirms that the analytic solution captures the dominant physics of transmission through a hydrodynamically escaping
atmosphere in the regime where the subsonic and step-function
approximations apply, while departures at strongly saturated
wavelengths motivate the more general hydrodynamic interpretation
discussed in section \ref{sec:hydro_interp}.
At several wavelengths, however, the analytic solution does not admit a real-valued Lambert-$W$
solution. A characteristic example is the Ly$\alpha$ line core. In the numerical model, the effective
radius reaches very large values, as expected for a strongly saturated resonance line. In contrast,
the analytic inversion fails because the real-domain condition in Eq.~\eqref{eq:Reff_W_minus1} is
violated, i.e.\ $z(\lambda)<-1/e$.  The numerical spectrum remains well defined in this
regime: it is the analytic inversion between the optical-depth criterion and the effective radius
that breaks down, not the underlying radiative-transfer calculation.

This connects directly to the saturation boundary derived in Sect.~\ref{sec:chi_boundary}. Wavelengths
for which Eq.~\eqref{eq:validity_boundary} is satisfied correspond to an opacity-limited regime in
which $R_{\rm eff}$ remains sensitive to $\dot M$ and other physical parameters. Wavelengths for which
the inequality is violated lie in a saturation-limited regime: the optical depth exceeds the adopted
threshold along all relevant grazing chords, and the mapping between absorption depth and mass loss is
no longer invertible. In this sense, saturation is not merely a gradual reduction in sensitivity, but a
sharp loss of invertibility that follows from the analytic structure of the solution.

\begin{figure}[ht!]
    \centering
    \includegraphics[scale=0.47, trim= 0cm 0cm 0cm 0cm, clip]{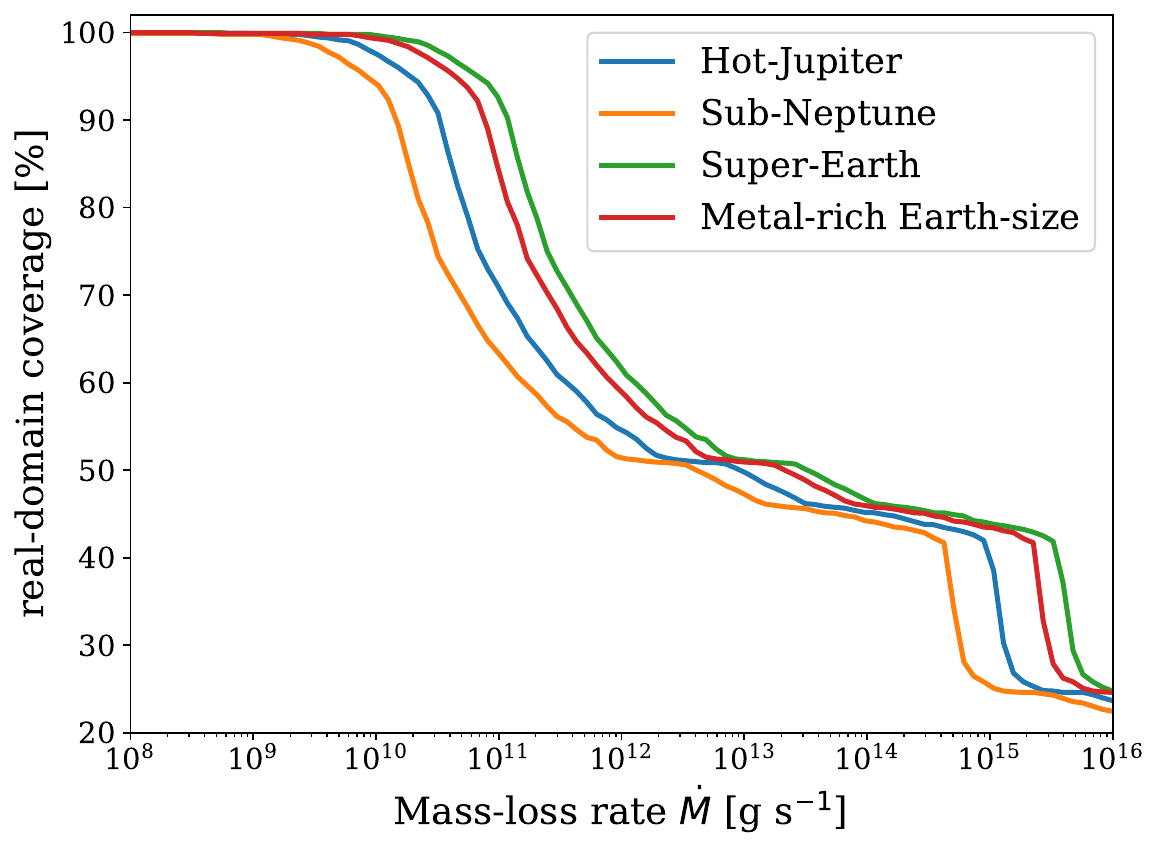}
    \caption{Analytic coverage as a function of wavelength, defined as the fraction of
wavelength points for which the Lambert-$W$ argument satisfies
$-1/e \le z(\lambda) < 0$. This condition ensures a real-valued effective
transit radius and therefore identifies the wavelength range where the
analytic inversion remains valid.
}
    \label{fig:fraction}
\end{figure}

Expressed in this way, the analytic Parker-wind transmission spectrum highlights that the observable
extent of an escaping atmosphere is determined by the interplay between opacity and mass flux, rather
than by pressure normalization alone. The sonic radius provides the natural geometric scale controlling
the extent of the optically thick region. In Fig.~\ref{fig:fraction} the fraction of wavelengths
in the ultraviolet that satisfy the real-domain condition $-1/e\le z(\lambda)<0$ are shown, i.e.\ the fraction of
wavelength points where the analytic inversion remains valid, across a range of mass-loss rates.

\subsection{Hydrodynamic interpretation of the saturation boundary beyond the steep-$\tau$ approximation}
\label{sec:hydro_interp}

The analytic derivation presented in the previous sections provides a
closed-form inversion between transmission depth and mass-loss rate
under the steep-$\tau$ approximation commonly used in analytic
transmission theory. The resulting saturation boundary therefore
identifies the regime in which this analytic inversion ceases to admit
a unique solution. In this section I examine the regime in which the steep-$\tau$ approximation fails, and show how the resulting saturation behavior can be interpreted in a more general hydrodynamic context. Even when the optical depth varies gradually with impact parameter (E.g. \citet{BallabioOwen2025}), the exact transmission
integral naturally separates opacity-limited and saturation-limited
regimes.

In the analytic inversion derived in the previous section, the transmission integral is dominated by a narrow annulus around $b_*$.  However, the fundamental observable is the exact transmission-area
integral  given in equation \ref{eq:Reff_def}  which does not require the steep-$\tau$ assumption. The transmission signal therefore naturally separates two regimes depending on the
magnitude of the optical depth.

\paragraph{Opacity-limited regime}
If the atmosphere is optically thin over most of the contributing
region,

\begin{equation}
\tau(b,\lambda) \ll 1 ,
\end{equation}

then $1-e^{-\tau}\simeq\tau$ and the transmission signal becomes
approximately proportional to the optical depth,

\begin{equation}
R_{\rm eff}^2 - R_0^2
\propto
\int \tau(b,\lambda)\, b\,db .
\label{eq:35}
\end{equation}

Since the optical depth scales linearly with the mass-loss rate,
$\tau \propto \sigma(\lambda)\dot{M}$, the observable absorption also
scales approximately with $\dot{M}$. In this regime the mapping between
transmission depth and mass-loss rate remains approximately invertible.

\paragraph{Saturation-limited regime}
If the inner region of the outflow becomes optically thick,

\begin{equation}
\tau(b,\lambda) \gg 1
\end{equation}

for $b<b_{\rm sat}$, the integrand approaches unity and the contribution
from this region becomes purely geometric,
\begin{equation}
\int_{R_0}^{b_{\rm sat}} b\,db
=
\frac{1}{2}(b_{\rm sat}^2-R_0^2).
\end{equation}

Increasing the column density further does not significantly increase
the absorption from this region. The transmission signal becomes
controlled by the projected size of the saturated region and by the
optically thin outer tail of the atmosphere rather than by the total
column density. In this regime the mapping between transmission depth
and mass-loss rate loses uniqueness.\\

We can now derive the scaling of the optical depth as a function  of the impact parameter, $b$, in the context of a general hydrodynamic outflow. Starting from the steady wind density profile
\begin{equation}
n(r)=\frac{\dot{M}}{4\pi \mu m_p r^2 v(r)},
\end{equation}
and inserting it into the chord optical depth integral, one obtains
\begin{equation}
\tau(b,\lambda)
=
\sigma(\lambda)\int n(r)\,ds
\;\sim\;
\frac{\sigma(\lambda)\dot{M}}{4\,\mu m_p}
\int \frac{ds}{r^2 v(r)}.
\end{equation}

Approximating the integral near the tangent point, where the
geometric kernel peaks (i.e.\ $r\sim b$) and $v(r)\sim v(b)$, yields. In particular, this step does not assume that the optical depth varies rapidly with impact parameter, but only that the dominant contribution to the integral arises from the vicinity of the tangent point due to geometric projection.
\begin{equation}
\tau(b,\lambda) \propto
\frac{\sigma(\lambda)\dot{M}}{b\,v(b)} .
\end{equation}
This scaling is consistent with the analytic condition derived in section \ref{sec:chi_boundary}, in which the product $\sigma(\lambda)\dot{M}$ acts as the fundamental control parameter determining whether the transmission signal lies in the opacity-limited or saturation-limited regime. 

A useful hydrodynamic estimate of the saturation scale can be obtained
by defining the impact parameter $b_{\rm sat}$ at which the optical
depth becomes of order unity,

\begin{equation}
\tau(b_{\rm sat},\lambda)\sim1 .
\end{equation}

Using the scaling above gives

\begin{equation}
b_{\rm sat}(\lambda)
\sim
\frac{\sigma(\lambda)\dot{M}}{v(b_{\rm sat})}.
\end{equation}

This relation shows that stronger transitions (larger
$\sigma(\lambda)$) and larger mass-loss rates extend the saturated
region to larger impact parameters, while faster winds reduce the
optical depth and shrink the saturated region.  The analytic saturation boundary derived above may therefore be interpreted as identifying the transition where the steep-$\tau$ inversion ceases to be valid and the transmission signal becomes dominated by the geometric extent of the optically thick region of the outflow. The hydrodynamic scaling above also clarifies the meaning of the
analytic saturation boundary derived in the previous sections. If the
onset of saturation is defined by the condition
$\tau(b_{\rm sat},\lambda)\sim 1$, then
$\tau(b,\lambda)\propto \sigma(\lambda)\dot{M}/[b\,v(b)]$ implies

\begin{equation}
\sigma(\lambda)\dot{M}
\sim
b_{\rm sat}\,v(b_{\rm sat}) .
\end{equation}

The product $\sigma(\lambda)\dot{M}$ therefore determines how far
outward the optically thick region extends along the line of sight.
The analytic saturation condition derived earlier may then be
interpreted as the local steep-$\tau$ realization of this more general
hydrodynamic transition: once the saturated region expands to the
characteristic transmission radius, the inversion between transit depth
and mass-loss rate ceases to be unique.

\section{Discussion}
\label{sec:discussion}

\subsection{Information content of transmission spectra in planetary winds}

The analytic framework developed in this work shows that transmission spectroscopy of hydrodynamically escaping atmospheres is fundamentally governed by the mathematical structure of the optical depth inversion problem. When the argument of the Lambert-$W$ function remains within its real-valued domain, the effective transit radius is a single-valued and monotonic function of the product $\sigma(\lambda)\dot M$, and the transmission spectrum is \emph{opacity-limited}. In this regime, variations in mass-loss rate, temperature, mean molecular weight, or absorber abundance lead to measurable changes in the transit depth, and the analytic solution accurately reproduces numerical radiative-transfer calculations.

Once the real-domain condition is violated, the inversion ceases to exist as a real-valued mapping. This defines a sharp transition to a \emph{saturation-limited} regime, in which the optical depth along all relevant grazing chords exceeds the reference threshold $\tau_\ast$. In this case, further increases in opacity or mass flux do not lead to a unique increase in the effective transit radius. This saturation regime reflects a failure of the analytic inversion, not of radiative transfer itself: fully numerical transmission spectra remain well defined and continuous in this limit. Transmission spectroscopy therefore loses its ability to uniquely constrain the atmospheric column density or the mass-loss rate at those wavelengths.

This behavior provides a natural explanation for the long-standing
result from numerical models that the cores of strong resonance lines
(e.g., Ly$\alpha$, H$\alpha$) often show weak sensitivity to $\dot M$
when they lie in the saturation-limited regime. In the analytic
framework presented here, this occurs when the line core lies beyond
the saturation boundary, such that the effective transit radius becomes
insensitive to further increases in opacity or mass flux.

For the He~\textsc{i}~1083~nm triplet,  previous studies have
shown that the observable absorption is often dominated by optically
thin regions of the flow, leading to a stronger dependence on $\dot M$ \citep{BallabioOwen2025, Linssen2023}. More generally, this highlights that  different spectral lines probe different regimes of the outflow depending on their opacity and excitation conditions.

From an observational perspective, this implies that transmission spectra should not be interpreted uniformly across wavelength. Instead, spectral regions should be classified according to whether they lie in the opacity-limited or saturation-limited regime. Only the former admit an approximately unique mapping between absorption depth and the underlying escape parameters within the present analytic framework.

\subsection{Geometric nature of saturated line cores}

In the saturation-limited regime, the Lambert-$W$ solution approaches the branch
point and the effective transit radius asymptotically converges to
\begin{equation}
R_{\rm eff} \rightarrow 2 r_s,
\end{equation}
where
\begin{equation}
r_s = \frac{G M_p}{2 c_s^2}
\end{equation}
is the sonic radius of the Parker wind. This saturation scale depends only on the
planetary gravitational potential and the atmospheric temperature through the sound
speed, and is independent of both the absorption cross section and the mass-loss
rate. The observable absorption depth in saturated line cores therefore reflects the
geometric extent of the optically thick region rather than the atmospheric column
density or mass flux. 

The existence of the saturation boundary follows from the analytic
structure of the optical-depth inversion and therefore arises
independently of the specific numerical implementation of radiative
transfer. Its precise location depends on the product
$\sigma(\lambda)\dot{M}$ and on the global thermodynamic properties of
the wind.  In real systems, the observable geometric extent may be truncated at smaller radii by ionization fronts, interaction with the stellar wind, or Roche-lobe effects.

This result demonstrates that the weak sensitivity of saturated line cores to $\dot M$ is not a numerical artifact or a gradual loss of signal, but a direct consequence of the non-invertibility of the optical-depth condition in an expanding atmosphere. Two planets with similar temperature and gravity but substantially different mass-loss rates can therefore exhibit nearly identical saturated line cores, provided that both lie beyond the analytic saturation boundary.

Consequently, strong transmission lines primarily constrain geometric properties of the upper atmosphere, such as the radial extent of the absorbing species and the termination altitude set by ionization or dissociation processes, whereas quantitative constraints on mass loss must rely on opacity-limited diagnostics.

Finally, it is worth noting that the expressions for the effective transit radius, Eq. \ref{eq:Reff_def}, \ref{eq:35}, implicitly assume that the optical depth decreases sufficiently rapidly at large impact parameters for the corresponding integrals to converge. However, in a Parker wind the density profile asymptotically approaches a $r^{-2}$ scaling in the supersonic regime, which implies that the chord optical depth does not decrease faster than $1/b^2$. As a result, the integral for the obscured area would formally diverge if extended to infinite radius. In practice, this divergence is avoided because real planetary outflows have a finite spatial extent. The escaping atmosphere is truncated by physical processes such as photoionization, interaction with the stellar wind, or confinement within the Roche lobe, which introduce an effective outer cutoff radius. The observable absorption is therefore dominated by the region where the optical depth transitions through unity, and the scaling relations derived here remain valid, as they are controlled by the inner regions of the flow rather than by the asymptotic behaviour at large radii.

\subsection{Connection to the escape parameter and hydrostatic structure}

The sonic radius provides a physically transparent link between transmission spectroscopy of winds and classical escape theory. Introducing the Jeans escape parameter evaluated at the planetary radius,
\begin{equation}
\lambda_0 \equiv \frac{G M_p \mu}{k T R_p},
\end{equation}
one obtains
\begin{equation}
\frac{r_s}{R_p} = \frac{\lambda_0}{2}.
\end{equation}

The saturation scale $R_{\rm eff}\simeq 2r_s$ therefore corresponds to a fixed fraction of the gravitational binding depth of the atmosphere. This highlights a fundamental difference with hydrostatic transmission spectra, in which the effective radius is anchored to an arbitrary reference pressure level. In planetary winds, by contrast, the observable extent of the atmosphere is controlled by the depth of the gravitational potential well and the thermal state of the gas.

\subsection{Interpretation in the energy-limited escape framework}

Although the analytic solution is expressed in terms of the mass-loss rate, it is instructive to reinterpret the results under the assumption of approximately energy-limited escape,
\begin{equation}
\dot M_{\rm EL} =
\frac{\eta \pi R_{\rm XUV}^3 F_{\rm XUV}}{G M_p K},
\end{equation}
where $\eta$ is the heating efficiency, $R_{\rm XUV}$ the effective absorption radius, $F_{\rm XUV}$ the stellar high-energy flux, and $K$ the Roche-lobe correction factor.

Substituting this expression into the dimensionless control parameter $\chi(\lambda)$
yields the scaling
\begin{equation}
\chi(\lambda) \propto 
\frac{\sigma(\lambda)\,\eta\,F_{\rm XUV}\,R_{\rm XUV}^3\,\sqrt{T}}
{(G M_p)^2\,\mu^{3/2}},
\end{equation}
up to numerical factors of order unity. This relation shows that, at fixed opacity, the transmission spectrum of a planetary wind is most sensitive to stellar irradiation and planetary mass,  and only weakly dependent on temperature, while retaining a stronger dependence on mean molecular weight through the sound-speed scaling. Within this framework, the analytic saturation boundary can be interpreted as a threshold in stellar forcing beyond which the subsonic region of the outflow becomes optically thick at the wavelengths considered. Strongly irradiated, low-gravity planets are therefore expected to exhibit saturated line cores over wide spectral intervals, whereas higher-gravity or weakly irradiated planets may remain in the opacity-limited regime even for intrinsically strong transitions.

\subsection{Implications for observations and retrievals}

The analytic results presented here provide a quantitative criterion for assessing the diagnostic power of different spectral tracers of atmospheric escape. Wavelength regions that satisfy the real-domain condition of the Lambert-$W$ solution admit a unique mapping between transmission depth and atmospheric parameters and are therefore suitable for quantitative mass-loss constraints. Regions that violate this condition lose unique sensitivity to the mass-loss rate and primarily probe the geometric extent of the absorbing atmosphere.

This distinction offers practical guidance for observational strategies
and retrieval analyses.  In particular,  weaker transitions and the wings of strong lines are more likely to remain in the opacity-limited regime,
where the transmission signal retains a direct sensitivity to the atmospheric column density and mass-loss rate, although weaker transitions may be more challenging to detect observationally due to their smaller absorption depth,
particularly for UV diagnostics. Strong lines, on the other hand, are often easier to detect and can
provide both line cores and wings, but their cores may enter the
saturation regime, in which case the observable absorption depth
primarily reflects the geometric extent of the escaping atmosphere
rather than the mass flux itself.

In practice, additional factors may further complicate the
interpretation of observations.  For example, Ly$\alpha$ line cores are
frequently obscured by interstellar absorption and the observable signal is typically inferred from the line wings, where interactions with the stellar environment may play an important role
(e.g. \citet{Owen2023}). Likewise, interpretation of the helium
1083~nm triplet depends on the thermodynamic structure of the outflow,
including the temperature, the H/He abundance ratio, and the fraction
of helium atoms in the metastable triplet state
(e.g. \citet{DosSantos2022}). Such effects introduce additional
degeneracies that can affect mass-loss estimates even when the
opacity-limited condition is satisfied.

\begin{acknowledgements}

The author acknowledges financial support from the Severo Ochoa grant CEX2021-001131-S funded by MCIN/AEI/10.13039/501100011033 and 
Ministerio de Ciencia e Innovación through the project PID2022-137241NB-C43.
\end{acknowledgements}

\bibliographystyle{aa} 
\bibliography{aa}

\appendix
\section{Numerical transmission model}
\label{app:dnumerical}
Fully numerical transmission spectra are computed using the Parker-wind density and velocity profiles. The atmospheric structure was described by a spherically symmetric, isothermal hydrodynamic outflow, for which the radial velocity $v(r)$ and number density $n(r)$ were obtained from the exact Parker solution. The profiles were computed from the planetary radius $R_{\rm p}$ up to $r_{\rm max}=10\,R_{\rm p}$ and were used directly in the radiative transfer calculation.

The wavelength-dependent extinction cross section of the atmosphere was constructed as a volume–mixing–ratio–weighted sum of the individual species cross sections,
\begin{equation}
\sigma_{\rm mix}(\lambda) = \sum_i x_i\,\sigma_i(\lambda),
\end{equation}
where $x_i$ denotes the volume mixing ratio of species $i$. The same mixing ratios as in the analytical models were adopted.

For a given wavelength $\lambda$, the chord optical depth at impact parameter $b$ was computed as
\begin{equation}
\tau_\lambda(b) = \int_{-\infty}^{+\infty}
n\!\left(r(s)\right)\,
\sigma_{\rm mix}\!\left[\lambda'(s)\right]\,
\mathrm{d}s,
\end{equation}
where $r(s)=\sqrt{b^2+s^2}$ is the radial distance along the line of sight. The Doppler-shifted wavelength $\lambda'(s)$ accounts for bulk hydrodynamic motion and is given by
\begin{equation}
\lambda'(s) = \lambda\left(1 - \frac{v_{\rm los}(s)}{c}\right),
\end{equation}
with $c$ the speed of light and $v_{\rm los}(s)=v(r)\,s/r$ the line-of-sight component of the radial wind velocity. This treatment naturally produces Doppler broadening due to the velocity gradient along the chord, as different regions of the atmosphere contribute at different projected velocities.

Thermal Doppler broadening was included by convolving the mixed cross section $\sigma_{\rm mix}(\lambda)$ with a Gaussian kernel in $\ln\lambda$, corresponding to a Maxwellian velocity distribution at the atmospheric temperature $T_0$. The fractional width of the kernel is $\sigma_{\ln\lambda}=v_{\rm th}/c$, where $v_{\rm th}=\sqrt{2k_{\rm B}T_0/m}$ is the thermal velocity of the absorbing species of mass $m$.

The wavelength-dependent transit depth was then obtained by integrating over all impact parameters,
\begin{equation}
\delta(\lambda) =
\frac{1}{R_\star^2}
\left[
R_{\rm p}^2 +
2\int_{R_{\rm p}}^{b_{\rm max}}
\left(1-e^{-\tau_\lambda(b)}\right)
b\,\mathrm{d}b
\right],
\end{equation}
where $R_\star$ is the stellar radius and $b_{\rm max}$ was chosen sufficiently large to enclose the optically thin upper atmosphere.

For resonance lines such as Ly$\alpha$, the numerical spectra were computed on a high-resolution wavelength grid ($\Delta\lambda \lesssim 10^{-3}$~nm) in order to properly resolve both thermal and bulk Doppler broadening. At coarser wavelength resolution, the impact of velocity broadening on the integrated transit depth is reduced, particularly for optically thick lines.

\end{document}